# Towards a Security Plane for 6G Ecosystems

Xavi Masip-Bruin[1], Eva Rodríguez[1], Admela Jukan[2], Panos Trakadas[3]

[1] Universitat Politècnica de Catalunya (UPC), CRAAX Lab, Vilanova i la Geltrú, 08800, Spain
[2] Technische Universität Carolo-Wilhelmina zu Braunschweig, Germany
[3] National and Kapodistrian University of Athens, Dpt. of Port Management and Shipping, Dirfies Messapies, Greece, 34400

*Abstract*—6G networks promise to be the proper technology to support a wide deployment of highly demanding services, satisfying key users-related aspects such as extremely high quality, and persistent communications. However, there is no service to support if the network is not reliable enough. In this direction, it is with no doubt that security guarantees become a must. Traditional security approaches have focused on providing specific and attack-tailored solutions that will not properly meet the uncertainties driven by a technology yet under development and showing an attack surface not completely identified either. In this positioning paper we propose a softwarized solution, defining a Security Plane built on a top of programmable and adaptable set of "live" Security Functions under a proactive strategy. In addition, in order to address the inaccuracies driven by the predictive models a pre-assessment scenario is also considered ensuring that no action will be deployed if not previously verified. Although more efforts are required to develop this initiative, we think that such a shift paradigm is the only way to face security provisioning challenges in 6G ecosystems.

## I. INTRODUCTION

THE evolution of mobile networks from 5G to 6G marks a transformative leap towards an ultra-connected world, where the convergence of communications, computing, and intelligence technologies redefines the boundaries of connectivity. As 5G continues to establish itself as a backbone of modern communications with enhanced bandwidth, reduced latency, and massive IoT integration, 6G has already emerged with the promise of revolutionizing the network ecosystem even further. 6G is expected not only to integrate advanced technologies, such as terahertz (THz) frequency band communications, but also the most comprehensive artificial intelligence (AI) capabilities, and sophisticated network slicing for services at unprecedented data rates in the order of terabits per second and near-zero latency [1]. 6G is envisioned to enable pervasive and ubiquitous connectivity, further extend the network and compute continuum, and combine public networks and non-public networks (NPN) to provide connectivity to underserved areas, aiming to extending network coverage and boost the performance end-to-end. The reals of 6G applications and services envision holographic communication and high-precision industrial automation, extended reality (XR), etc., laying the foundations for a deeply interconnected and intelligent digital society of the future.

Considering the scale and rapid nature of this network evolution, it has become both technologically and conceptually rather challenging to safeguard security and privacy of the system and its users, such as when protecting sensitive user or critical data, securing ultra-low latency communication, detecting novel cyber threats, and ensuring the integrity of the network infrastructure. Massive connectivity intrinsic to 6G networks increases the attack surface, and urgently necessitates innovative approaches to secure a multitude of interconnected devices and applications. In addition, the integration of artificial intelligence and the related machine learning algorithms in 6G networks introduces concerns related to the privacy of user data and the potential biases and misuse of AI algorithms. The entanglement of devices and technologies in a cloud continuum which ranges from the hyperscale cloud platforms to the very far edge requires full coordination, increasing the complexity of the management of networks, while introducing heightened security challenges [2]. The broad scope of 6G network that integrates aerial and satellite networks (Non-Terrestrial Networks, NTN), Non-Public Networks (NPN) and Public Network Integrated NPNs (PNI-NPN) furthermore amplify security and privacy challenges.

In this highly evolving scenario, key contextual constraints and requirements must be properly addressed towards designing a secure 6G ecosystem. Next, we summarize them in three main concepts, and highlight a potential pathway to sort them out.

First, a 6G ecosystem becomes a highly distributed, dynamic and continuously evolving scenario where many different technologies and solutions emerge as they are developed to improve 6G network services delivery. This results in several yet active efforts to design what the 6G architecture will be. This uncertainty drives the need for a softwarized strategy that may adapt to any known and yet non-foreseen security threat.

Second, as technology advances, the 6G ecosystem threat landscape continues to evolve towards a not yet completely identified attack surface. This continuous growth, where interdependencies between architectural components and cascading effects pave the way to new attacks, is enough to guarantee that a static and reactive approach based on security solutions addressing specific security threats will not be valid anymore. Instead, two key paths should be adopted. On one hand, attacks should be modelled to infer as much information as possible in order to be able to "characterize" new attacks, potentially contributing to existing frameworks, such as MITRE FiGHT [3] with specific 6G attacks, supported by a cooperative scenario where datasets may be shared to create behavioural attack models, for example, using GenAI to develop new attacks or augment the required data. On the other hand, push for a proactive solution responsible for predicting attacks and deciding preventive actions to be deployed.

And third, services deployed in a 6G ecosystem, such as those mentioned before in this section, are expected to be extremely demanding in terms of quality, what indeed is strongly affected by the network reliability. Considering security breaches and the effects derived in terms of service disruption, a significant contributor to reliability, the question is not only what the attack is, but what the impact the attack would have in the infrastructure will be and also important what the impact of the mitigation action to be deployed will be as well. A potentially strategical approach to handle this impact analysis context leverages a pre-assessment scenario that may on one hand analyse attacks impact and on the other hand guarantee that *no* action is deployed *wi*thout being preliminary *te*sted, defined as the NOWIT principle).

The key research question here is what kind of solution should be designed that can endow a network ecosystem like 6G not completely defined yet, with strong security guarantees, considering the dynamics, heterogeneity, as well as the large set of involved baseline tools and network technologies, assuming a friendly and collaborative management to facilitate systems interoperation. In this paper, we position a paradigm shift, moving from traditional isolated and usually proprietary solutions for security provisioning towards an integrated system, defined in terms of a Security Plane (SP), mirroring behaviour of well-known Data or M&O Plans. This SP builds the whole security provisioning recipe upon three key fundamental ingredients, namely, softwarization, proactivity and the NOWIT principle, further elaborated in this manuscript, that must also accommodate the specific needs the three separated 6G domains would pose, devices (6G UEs), Access Networks (6G NB, RAN sharing) and Core Network (including both 5G legacy and new NFs).

The rest of the paper is hence organized as follows. Section II reviews the key challenges in 6G ecosystems. Section III introduces the key concepts behind the proposed Security Plane idea that are further elaborated in Section IV. Finally, Section V concludes the paper.

## II. Security in 6G Ecosystems: Key Challenges

It is with no doubt that the three contextual constraints and requirements defined in previous section will drive relevant research challenges. Considering the fact that, as of today, the 6G ecosystem is characterized by a high level of uncertainty, inferred from the lack of a widely adopted architecture and the continuous adoption of new technologies, in this paper we propose to link the envisioned challenges to a specific short set of Technology Trends, fueling the next generation of network and communication systems.

Network disaggregation, usually referred to as the separation of hardware and software components, enables a modular network setup where functions and services can run independently across different physical and virtual assets. However, network disaggregation introduces new challenges, such as ensuring interoperability, maintaining consistent performance levels, and securing a more complex network architecture that could be vulnerable to cyber threats.

The integration of Non-terrestrial Networks (NTNs), which include satellite networks and high-altitude platforms, into the 6G ecosystem presents unprecedented possibilities for ubiquitous global connectivity. However, the extended reach and the heterogeneous nature of these networks raise significant security and privacy challenges, for example to ensure coherent data encryption and secure handoffs between terrestrial and non-terrestrial segments or to guarantee privacy over NTNs.

Private networks, referred to as Non-Public Networks (NPNs) within the 3rd Generation Partnership Project (3GPP) [4] become a fundamental element for communication in the era of Industry 4.0. The integration of NPN management with the public network management (NPI-NPN) introduces security challenges, such as authentication, encryption as well as privacy of private network generated data.

The network slicing concept and technology already introduced in 5G, enables more service providers to offer customized networks flexibly with different functionalities for either diverse services or to serve groups of users with specific service requirements. However, the fact that common infrastructure and resources are shared amongst network slices needs to be taken into consideration for 5G systems pursuing the security-by-design principle as laid out in 3GPP Standards [5].

One key innovation in the 6G era will be the new design of the system architecture to support the extreme cloud continuum, where network functions from different 6G network segments can be composited flexibly and dynamically based on service needs in diverse cloud environments. In this highly distributed system, solutions aimed at managing security concerns resulting from the correlation between multiple domains, the execution of computationally intensive applications smartly distributed in the continuum, the varying levels of security controls implemented across different domains and the utilization of barely protected edge systems to trigger powerful attacks are critical.

With the advent of 6G, hardware and software will become more independent (decoupled), allowing for a modular and separated software-driven Radio Access Networks (RAN) built with cloud-based microservices fostered by architectures like O-RAN [6]. While these architectural changes offer advantages, they also raise complex security and privacy challenges basically related to ensuring a secure interoperation and interoperability within the envisioned multi-domain ecosystem that need to be addressed.

In this context of continuous technology evolution, existing efforts on two specific aspects deserve specific attention.

First, the envisioned 6G threat landscape along with enabling technologies, such as AI/ML, network openness, open-source software, etc. increase the attack surface and pose new threats [7], beyond considering those traditionally running in mobile and in general wireless networks. Indeed, even though 6G is a natural extension of 5G, when related to security attacks this assessment can only be taken as a seed context for the definition of attacks, as the inclusion of new functionalities and technologies in 6G ecosystems is opening up the attack surface to a yet partially unforeseeable dimension. This is not only due to the emerging technologies (AI as a means to attack the 6G network, massively distributed denial-of-service, etc.) but also due to the newly introduced Network Functions, both on the Radio Access (e.g. rApps, xApps and dApps, Conflict Management, real-time RIC) and on the Core part of the 6G network (NWDAF and MDAF,

CAPIF, integration of non-terrestrial networks, etc.), introducing new types of attacks (e.g., driven by the upcoming, already under study, applications of Generative AI both for malicious intents). Thus, it is crucial to characterize and provide defense countermeasures to protect against these threats. Reconfigurable intelligent surfaces (RIS) are considered promising for 6G networks but they are vulnerable to various security threats and risks. The security obstacles faced by RIS-empowered 6G networks encompass a range of attacks targeting diverse applications, such as millimeter wave communication, Internet of Things (IoT) networks, and NTNs [8]. The integration of AI with a 6G network brings an opportunity for innovation and solutions to different attacks. However, it also poses different challenges in terms of network security and privacy [9], including for example: i) the implementation of automated and distributed services; ii) AI training manipulated by injecting fake signals.; iii) other malicious attacks can eavesdrop through the interception of wireless channels, including DoS attacks, spoofing, and malicious data injection. MITRE ATLAS framework [10], an international standard developed for ML-based attack model mapping, contains a set of tactics and techniques used by the adversary to gain access to the ML model and to accomplish a specific objective. To address the aforementioned areas of 6G security and to mitigate the impact of an attack, there is a need to address the attack distinctively along with providing the information related to the attack vector. In the existing literature, some efforts have been made to identify different types of threats for 6G architecture. For example, authors in [11] identified the threats in the physical layer, connection (network) layer, and service (application) layer and elaborated on the privacy and security concerns and their respective solutions in enabling technologies of 6G infrastructure. In [12] a deep learning model for the detection of attacks occurring on edge computing in a 6G network is proposed.

Second, the application of AI is gaining importance in 6G to design proactive strategies for security management. Unlike 5G networks, where security solutions across all devices and base stations are configured with universal settings for certain types of attacks, it is apparent that such an approach cannot be applied in 6G networks. In this case, the high diversity in service and application provision, connected devices and associated protocols in heterogeneous networks, as well as the various physical layer encoding and transmission techniques, render a highly complex environment with different requirements and settings. Since each scenario may have unique security requirements and energy availability, the selection and configuration of security strategies need to be optimized for 6G networks in an adaptive and dynamic manner. Due to the dynamicity, sophistication and advancements in the security threats, security administrators are facing unprecedented challenges when managing and protecting their systems. In [13], authors proposed an optimization framework to address the identified challenges in 6G networks. In [14], the authors analyzed various potential new threats caused by the introduction of new technologies related to the usage of open-source tools and frameworks for 6G network deployment and present possible mitigation strategies to address these threats. In general, AI solutions are based on centralized data collection [15], an approach that imposes several limitations regarding the user privacy exposure, while it is not aligned with the distributed architecture of the 6G networks. In [16] the use of Generative Adversarial Networks (GANs) is explored to simulate intrusions and malware for improving its detection, and to fuel defense against different attack methods.

Thus, in summary, current state-of-the-art is evidencing the fact that 6G ecosystems bring in a yet not completely defined network scenario, that must support the execution of highly demanding intensive services, in a context of many uncertainties. In this widely open scenario, this paper proposes a softwarized approach turning into the concept of a Security Plane for 6G.

III. 6G SECURITY PLANE: CONCEPTUAL APPROACH

This paper positions a paradigm shift in the rules for governing security provisioning in 6G networks. The key fuel of the proposed shift is driven by moving away from isolated, proprietary, reactive, static and attack-tailored approaches that will hardly accommodate the specific demands of a largely dynamic, heterogeneous and yet unknown 6G ecosystem, towards a softwarized approach. Indeed, security is typically handled under a reactive approach where specific attack-tailored solutions are deployed after considering the potential attacks to come supported by a previous system vulnerabilities check. Once an incident is detected, mitigation actions are deployed to reduce the impact of the already received attack. In fact, forensic strategies are mainly used to select the mitigation action best reducing the impact of the attack in the system, after analyzing past experiences using different mitigation solutions. This scenario may work in systems not requiring real/right-time processing or for contexts where business continuity is not a major objective. However, when addressing high speed 6G ecosystems that would support the deployment of infrastructures with real-time needs (immersive apps, health, manufacturing processes, autonomous vehicles, V2X, et.), reactivity in front of an attack (i.e., responding to an attack in reaction time) becomes a "must not", as it explicitly assumes the system and the running services are already disturbed.

In a different strategy, the proposed softwarized approach leverages a modular, adaptable and proactive strategy, coined as a Security Plane for 6G ecosystems, open to accommodate new sources of data, adapt to new 6G technologies, deal with new attacks, and to be used by third parties to deploy new ad-hoc defined Security Functions (SFs). Inferred from the NF paradigm widely adopted by 5G systems and consequently easing the transit from 5G to 6G ecosystems, SFs are the key component to design the SP. Table 1 shows the key innovative differences brought in by the proposed Security Plane strategy versus traditional solutions.

**Table 1**. Key differences contributed by the Security Plane approach

| Traditional | Security Plane |
|---|---|
| Locked, HW-oriented | Softwarized approach |
| Attack specific | Open, attack-"free" |
| Siloed & proprietary | Adaptable & Customizable |
| Limited to known attacks | Open to new attacks |
| Low interoperability, no sharing | Collaborative approach |
| Reactivity | Proactivity & Reactivity |
| Uncertainty & no impact analysis | NOWIT principle |
| Significant time to recover | Reduced recovery time |

Three main architectural elements are conceptually considered in the design of the proposed SP.

The first architectural element refers to the SF composition process. Indeed, the process to build a SF is supported by three main conceptual components, as shown in Fig. 1:

- The DOTs: DOTs are defined as security assets, including tools, algorithms, or any atomic security elements that may contribute to deliver a security capability.
- Security Functions (SF): SFs are independent and modular blocks of software offering specific security functionalities, in the form of microservices in order to be easily extendable and connected among them. This will be delivered to any architectural 6G Blueprint as enablers easing the provision of security and privacy in 6G ecosystems. A SF may be created as either a composition of DOTs or a combination of SFs.
- SF Composer: Supported by the set of requirements to be addressed, the SF Composer will "*connect the dots*" by dynamically and smartly picking the set of DOTs to build a SF that will properly manage an attack.

Although it is an ongoing effort, while the design of the SFs composing process must be wide enough to accommodate existing but also new attacks, the set of DOTS must also be enough to support the required orchestration process. Fig. 2 shows a preliminary list of potential envisioned SFs and DOTS, where DOTS are split into five categories, guaranteeing that at least one component within each category can interface with other components within the rest of categories. Moreover, for the sake of illustration, two examples of SF composing are described next.

1) **Automated Pentesting**

The Automated Pentesting SF aims to perform a continuous assessment of the 6G infrastructure to identify potential vulnerabilities, assess their impact and provide mitigation actions, being adapted to the dynamic 6G environments by continuously monitoring and testing for vulnerabilities in real-time, ensuring that security controls remain effective even as the 6G infrastructure evolves and scales to support new applications and services. Pentesting activities are costly in terms of both, time and money, and require highly specialized profiles, making them difficult to replicate frequently over time. To address this challenge, this SF will use AI, not only to refine the process, but also to mirror the depth of understanding, innovation, and versatility characteristic of human expertise, thus simulating the behavior of a human pentester, making informed decisions and interacting dynamically with complex and uncertain environments.

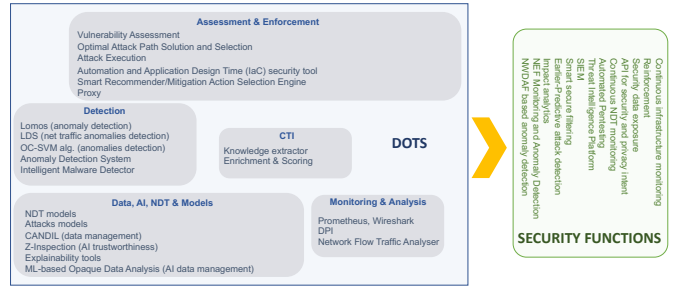

**Fig. 2.** Example of tentative DOTS and SFs

This SF includes three different DOTs. The related DOTs are: i) Vulnerability Assessment: this DOT will be responsible for scanning the infrastructure using different tools and methods, uncovering potential vulnerabilities, enriching the knowledge about them using different information sources and assessing their impact and relevance on the target infrastructure (this DOT will be assessing for instance the security of VNFs, to ensure they are isolated and prevent their unauthorized access, and/or the security of SDN controllers and switches ensuring the integrity and availability of the network; ii) Optimal Attack Path Solution and Selection: this DOT will make use of Reinforcement Learning (RL) and Markov Decision Processes (MDPs) to determine the optimal strategies for navigating through the penetration testing process efficiently, and; iii) Attack Execution: this DOT can execute an attack plan, containing a sequence of actions to be taken against the target infrastructure. The results of the execution of each action in the attack plan are analyzed to determine whether it is successful or not. In case the action fails, this means that the next state in the attack graph is not reachable. This information is used in an iterative process to determine the new best attack plan.

2) **Predictive Attack detection**

The Predictive Attack detection SF starts capturing network-flow traces using a network data collection tool (e.g., tcpdump) in the network infrastructure and saving the data in a PCAP format. The next step is to reduce the raw data intended to utilize only useful data for the later execution of ML algorithms. To this end, the PCAP files are processed by a network flow traffic analyzer tool (e.g., CIC-FLOW METER), extracting several useful network features that are stored in a CSV file, later converted into a JSON format. This data is simultaneously used for predictive detection combining unsupervised and supervised ML models: i) unsupervised components are used to filter between normal traffic and anomalies, and then possible attacks might be inferred, using specific ML algorithms (e.g., OC-SVM algorithms, One–Class Support Vector Machine); ii) supervised algorithms (e.g., Random Forest) are executed in parallel to classify the detected anomalies as different known attacks. These two types of algorithms work together in different combinations, phases, to classify known attacks as well as to detect zero-day attacks.

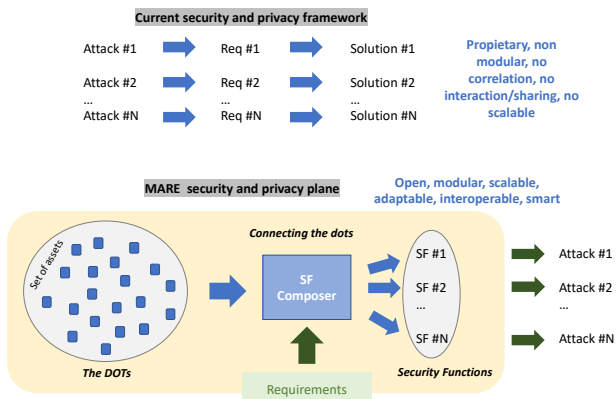

**Fig. 1.** Connecting the DOTS towards a SF

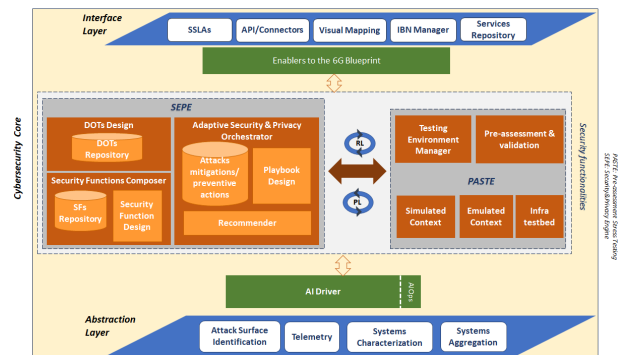

**Fig. 3.** Security Plane functional architecture

The second architectural element in the SP, refers to the fact that, in order to facilitate the deployment of reliable 6G systems, the proposed SFs must be properly orchestrated. The proposed SP should then include an orchestrator responsible for dynamically orchestrating the proposed SFs facilitating the deployment of reactive and proactive approaches, properly decided to best suit the expected needs and requirements.

Finally, as a third architectural element, the SP proposes a pre-assessment strategy, where preventive actions to be taken to actuate in front of a previously predicted attack, are analyzed to measure the potential impact these actions may have on the real infrastructure. The key concept behind this idea is the so-coined NOWIT paradigm (not without being tested). The main rationale resides in the fact that predictive approaches are based on estimations, what consequently inherits a significant uncertainty in the level of accuracy. Thus, in the worst scenario, the deployment of a preventive action might be worse than the attack it's supposed to protect to. In order to minimize this negative effect, the SP includes a functionality, referred to as pre-assessment that will be responsible to execute the proposed preventive measure in a sandbox context, adequately mirroring the network infrastructure to protect. Certainly, in order to make this approach realistic, the prediction should be done early in time to give the way to run the pre-assessment tests and the Digital Twin in the Sandbox must accurately represent the real infrastructure and dynamically adopt any change and modification.

## IV. Security Plane Modular Architecture

So far, in the previous section the conceptual roots of the proposed SP have been outlined. However, we want to go deeper in the concept description, proposing a preliminary functional architecture. The architectural layout proposed for the envisioned Security Plane is drawn in terms of three layers (see Fig. 3), each identifying a specific set of functionalities and also including potential tools to support them. These layers are described below.

### A. Abstraction Layer

The Abstraction Layer is responsible for: i) feeding the functional modules in the security plane in charge of defining DOTs and composing SFs, with monitoring data collected from the systems within the infrastructure used to obtain knowledge about the status of the systems; ii) defining the whole set of attacks to come, i.e., the attack surface; iii) define mechanisms for systems characterization and aggrupation, probably aligned to the attacks modelling, in order to map the real infrastructure into the PASTE functions. Information about the infrastructure is collected through a telemetry module that will leverage existing monitoring and telemetry tools (e.g., WireShark, Prometheus) to properly gather the required information. It is worth noting that the selection of this information is also challenging considering the multi-domain E2E scenario envisioned in 6G putting together access, transport and core technologies. The collected data should be properly classified and grouped to accurately characterize the set of elements in the overall E2E 6G ecosystem. This becomes a key aspect to maximize the correctness of the SF definition.

On the other hand, this layer will define the potential attacks to come up in a 6G ecosystem, hence defining the entire 6G attack surface. This task is also challenging as the 6G ecosystem is not yet completely defined. The outcome of this layer, i.e., attacks models and data (including infrastructure and user), will be forwarded to the Cybersecurity Core, to be properly and smartly (AI Driver) processed.

### B. Cybersecurity Core

The Cybersecurity core is the pivotal component in the envisioned Security Plane. It starts with the definition of the set of DOTs and delivers the orchestrated workflow of Security Functions for security provisioning. The envisioned Cybersecurity Core is however much more than that. Indeed, the envisioned functionalities allocated into this module are split into two main functional blocks (Fig. 1.3), Security and Privacy Engine (SEPE) and Pre-assessment Stress Testing (PASTE). Broadly speaking, SEPE is responsible for delivering the correct strategy (specific Security Function or in some cases a set of SFs) for security provisioning, and PASTE is responsible for both preliminary assessing the quality of the actions (i.e., impact, performance) to be taken based on the SEPE outcome and predicting potential attacks to come. It is worth mentioning that both modules work under a very close interaction to optimally deliver Cybersecurity Core capabilities.

SEPE works on two different dimensions. On the one hand it keeps working on defining and collecting DOTs and creating new composed Security Functions for each specific security attack to guarantee maximum security and privacy for existing and new attacks. These SFs are stored in a services repository located in the Interface Layer, and may be used internally and also offered to third parties. SEPE starts identifying a set of security assets, referred to as DOTs, that will be the atomic security elements (e.g., tools, technologies, algorithms, models, etc.) used to create the security functions (SFs). On the other hand, it runs both detection and prediction functions (composed by monitoring, analyzing, detecting and predicting functions), whose outcome will trigger the execution of a SF, either stored in a repository or after being composed using the proper DOTs (or a combination of SFs). The detection process is based on the monitoring data collected from the Abstraction Layer or by any other source of data that may contribute to get more knowledge on the real infrastructure status. The collected data is analyzed in the SEPE (detection tool) and the proper SF is selected or generated in the composer. SEPE can also manage the deployment and reconfigurability or extension of the attestation capabilities and trusted extensions of the network nodes. Differently, the prediction process will be deployed by the PASTE component where the Network Digital Twin (NDT) setting the Sandbox and the different attack models will set predictive strategies to infer the chances for an attack to come. Policies, some inferred from the Security SLA(SSLA), should be defined to analyze when these chances are large enough to launch a preventive action.

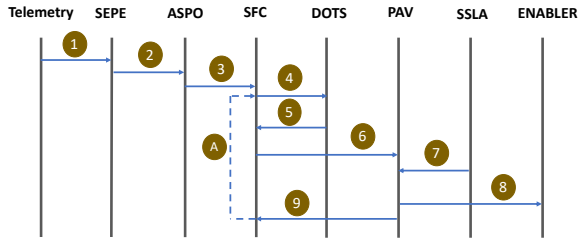

**Fig. 4.** Detection Workflow

Next, two illustrative workflows are introduced to illustrate the flow of actions in the proposed architecture. Fig. 4 describes the process deployed when an attack is detected in the real infrastructure.

The process starts in the telemetry component that collects data to feed the SEPE module. This data reaches out the SEPE (step 1), particularly the detection function (part of a security function already running), that will be responsible for analyzing the data and detecting a potential attack. If so, the Adaptive Security & Privacy Orchestrator (ASPO) will receive the kind of attack (step 2) and based on any potential rule of policy to meet, will decide the orchestration strategy, referred to as an orchestration playbook (for example it might be a mitigation-based approach aimed at guaranteeing business continuity), and will ask the Security Functions Composer (SFC) (step 3) to instantiate a Security Function particularly addressing the required needs. The SFC will either look at the Services Repository for any existing SF already covering the needs of the decided orchestration playbook (not shown in the workflow), or create a new Security Function. To this end, the SFC will ask the DOTs component (step 4) for the proper DOTs (step 5) and will create the SF that will be delivered to the Pre-assessment & Validation component (PAV) (step 6). The PAV, according to some policies to meet (Step 7) will decide if the proposed SF is sufficient to guarantee the demands set in the orchestration playbook (i.e., business continuity). This process may run on either all or some of the different validation contexts included in PASTE module (i.e., simulation, emulation, real infra), according to a policy to be defined (might be also part of the orchestration playbook). If it does (step 8) the SF is forwarded as an Enabler to the 6G Architecture. Otherwise, (step A) the SFC is contacted again to define a better SF. This process may end up in a never-ending loop, therefore, a predetermined policy will be also defined for a timely and efficient response to the attack.

On the other hand, Fig. 5 depicts a preliminary prediction workflow. In this scenario, the process is triggered by PASTE module. The Pre-assessment & Validation (PAV) component estimates with a certain probability that an attack is about to come (Step 1). Upon receiving the policies defining the potential impact this attack will have on the infrastructure to be protected (Step 2), the PAV launches a query to the ASPO to generate an orchestration playbook (step 3). One generated, the ASPO asks the SFC to define the preventive Security Function (step 4). If the SF already exists in the Service Repository the SF will pick this SF (not shown in the workflow), otherwise it will ask the DOTs components (step 5) for the set of DOTs (step 6). With the received DOTs, the SFC will compose a SF that will be sent to the PAV (Step 7) to analyze the impact this preventive action may have on the infrastructure. It is important to notice that although the steps in both workflows are similar, the activity to be done in the components is different. The PAV will estimate the impact using any of the existing validation contexts. Then, based on the policy identified in SSLA (step 8), PAV will decide if it is worth to deploy the SF (step 9) or not. If it is not, PAV reaches out the SFC to ask for composing a different SF (step A), starting the process (B). As in the detection workflow, this back process may be adjusted or even removed depending on the attack, the policy implemented, the business and/or infra to protect and the specific user demanding the protection.

These two workflows clearly describe what the real process for the proposed SP will be.

*C. Interface Layer*

The Interface Layer is responsible for easing the connections and visibility of the proposed security plane. This rationale is supported by facilitating: i) a comprehensive knowledge of the different actions and functions the Security Plane will execute, through a Visual Mapping component; ii) an smooth interaction with potential users through and Intent-Based Networking approach; iii) interoperability with external sources and third parties, through a set of API/connectors; iv) the adoption, tuning and definition of security policies that will be key elements feeding several decision making processes, such as for example the orchestration playbook created by the ASPO component, and; v) the deployment of an open, modular and adaptable security strategy built on top of a set of security functions, stored in a Security Services Repository, that can be (according to some rules to be defined) used, extended and modified by third parties willing to contribute to the SF definition.

The IBN Manager, see Fig.6, will translate the customers' SSLAs into intents in appropriate data model formats. A key feature will be introducing privacy intents as fields in the provided templates, complementary to service level

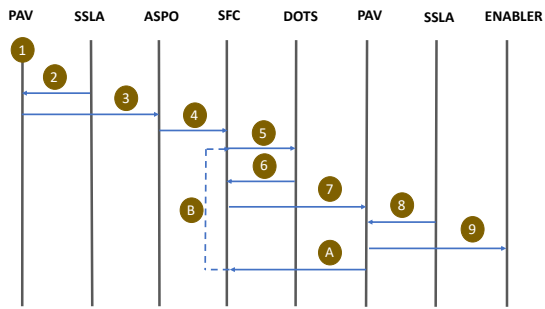

**Fig. 5.** Prediction Workflow

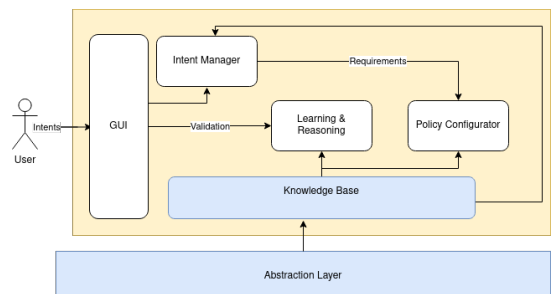

**Fig. 6.** Intent-based Network components

requirements. More specifically, this module acts as a bridge between user input and configured policies. It uses machine learning techniques to understand the user's intent (what they're trying to achieve) from their high-level text input. The GUI is a user-friendly dashboard designed to capture this input. The intent manager then takes over, storing and parsing the received intents and allows for full management of intents (create, read, update, and delete). Each intent has a defined structure that specifies its requirements and options. The intent manager identifies the most suitable policy from the knowledge base by matching the intent's requirements with existing policies. It ensures compatibility and consistency between user intent and system policies. Administrators are empowered by AI and machine learning in this system. By analyzing historical data on executions and decisions (gathered from the Abstraction Layer), the system recommends the most effective policy for future utilizations

The SSLA component is responsible for the definition of the scope and standards of security services offered and should stablish clear and measurable security performance criteria expressed in a machine-readable form that will allow automation of the security strategies and functions. The SSLA will provide for: i) Security Objectives and Metrics (i.e., confidentiality, integrity, availability and privacy levels); ii) Incident Response limits; iii) Data Protection (i.e., types of data encryption); iv) Remediation plans (automatic mitigation of detected anomalies, real-time alerting, corrective actions). The starting point for the definition of SSLA will be the GSMA Slice Template [16]. The proposed Generic Slice Template will be extended so it can accommodate SSLA specification along with all the other deployment specific provisions. The file is then processed by the IBN Manager in order to compose, orchestrate and provision the required security strategies and functions corresponding to the intent and security specifications.

## V. Conclusion

This paper proposes a paradigm shift in how security guarantees are provisioned in an extremely demanding and uncertain network ecosystem. The paper paves the way to continue to work on extending the network function paradigm to be adopted within the security field, building the concept of the Security Plane.

Recognizing the focus of the paper on positioning the SP concept, it also fuels new research avenues, including how DOTS must be composed to dynamically set the SF best suiting the incident to address, how different SFs may be orchestrated to set the proper proactive or reactive workflow, how these workflows may handle an e2e ecosystem covering many distinct network segments, how the inaccuracies driven by predictive models may be minimized, how may the impact of a preventive action may be identified, or how the NOWIT principle may be efficiently deployed. Moreover, notable challenges are also opened on how AI strategies, either ML-based to support the SF composition process, Gen AI-based to generate attacks diversity, or AI agents to facilitate autonomous and automated management may be applied. Also worth to mention the need for analyzing the deployment strategy best suiting the wide, extremely dynamic and heterogeneous 6G ecosystem.

Finally, the next step in this work is to start developing the SP concept, by analyzing the attack surface in 6G ecosystems, identifying the set of DOTS, defining the composing and orchestration strategies, setting the sandbox scenario and defining the right policies and compliance rules to support a much secure 6G ecosystem.

### Acknowledgment

This work has been partially supported by the HE MARE project, funded by European Commission with Grant Number 101191436, and for UPC authors by the Spanish Ministry of Science and Innovation under grant PID2024-156150OB-I00, funded by MCIN/AEI/10.13039/501100011033 and by ERDF A way of making Europe, as well as by the Catalan Government under contract 2021 641 SGR 00326.

### References

[1] E. K. Hong et al., "6G R&D vision: Requirements and candidate technologies," in Journal of Communications and Networks, vol. 24, no. 2, pp. 232-245, April 2022, doi: 10.23919/JCN.2022.000015
[2] M.Ishtiaq, et al. "Edge computing in IoT: A 6g perspective," 2022, https://arxiv.org/abs/2111.08943
[3] https://fight.mitre.org
[4] https://www.3gpp.org
[5] TR33.811(Rel15), TR33.326 (Rel17), TS33.501(Rel18)
[6] ORAN Alliance. O-RAN White Papers and Resources. https://www.o-ran.org/resources, 2024
[7] Meng, Rui, et al. "Multi-Dimensional Fingerprints-based Multi-Attacker Detection for 6G Systems." *IEEE Internet of Things Journal* (2023).
[8] Ouyang, Ye, Xiaozhou Ye, and Xidong Wang. "6G Network Operation Support System." *arXiv preprint arXiv:2307.09045* (2023)
[9] Siriwardhana, Yushan, et al. "AI and 6G security: Opportunities and challenges." *2021 Joint European Conference on Networks and Communications & 6G Summit (EuCNC/6G Summit)*. IEEE, 2021
[10] https://atlas.mitre.org
[11] Nguyen, Van-Linh, et al. "Security and privacy for 6G: A survey on prospective technologies and challenges.", *IEEE Communications Surveys & Tutorials*, 23.4 (2021): 2384-2428
[12] Abdelrahim, Elsaid M. "A Novel Data Offloading with Deep Learning Enabled Cyberattack Detection Model for Edge Computing in 6G Networks.", *AI-Enabled 6G Networks and Applications*, (2023): 17-33
[13] S. Shen, et al., "Adaptive and Dynamic Security in AI-Empowered 6G: From an Energy Efficiency Perspective," in IEEE Communications Standards Magazine, vol. 5-3, 2021, 10.1109/MCOMSTD.101.2000090.
[14] D. Je, et al., "Toward 6G Security: Technology Trends, Threats, and Solutions," in IEEE Communications Standards Magazine, vol. 5, no. 3, pp. 64-71, September 2021, doi: 10.1109/MCOMSTD.011.2000065.
[15] Nguyen, D. C., Ding, M., Pathirana, P. N., Seneviratne, A., Li, J., & Poor, H. V. (2021). Federated learning for internet of things: A comprehensive survey. IEEE Communications Surveys & Tutorials
[16] GSM Association Official Document NG.116 - Generic Network Slice Template, Version 8.0, 27, 2023